\documentclass{sf2a-conf2024}
\usepackage{graphicx}
\usepackage{hyperref}
\usepackage[]{natbib}  
\usepackage{epstopdf}

\def\BibTeX{{\rm B\kern-.05em{\sc i\kern-.025em b}\kern-.08em
    T\kern-.1667em\lower.7ex\hbox{E}\kern-.125emX}}
\bibpunct{(}{)}{;}{a}{}{,}  

\begin{document}

\TitreGlobal{SF2A 2024}


\title{The impact of differential rotation on the stochastic \\ excitation of acoustic modes in solar-like stars}

\runningtitle{The impact of differential rotation on the stochastic excitation of acoustic modes in solar-like stars}

\author{G. Biscarrat $^{1,}$}\address{Université PSL, Paris, 75006, France}\address{Université Paris-Saclay, Université Paris Cité, CEA, CNRS, AIM, Gif-sur-Yvette, F-91191, France}
\author{L. Bessila $^{2}$}
\author{S. Mathis$^{2}$}




\setcounter{page}{237}


\maketitle


\begin{abstract}
We model the stochastic excitation of acoustic modes in solar-like pulsators taking into account the action of differential rotation. We derive the theoretical formalism for the stochastic excitation with differential rotation, and make use of rotating convection Mixing-Length Theory to assess how the convective velocity is modified by rotation. Finally, we use the stellar structure and evolution code MESA combined with the stellar pulsation code GYRE to compute acoustic modes amplitudes.\end{abstract}



\section{Introduction}

Acoustic modes are stochastically excited by turbulent convection in solar-like stars' envelopes \citep[e.g.][]{samadi_excitation_2001}. Recent observations using the results from \textit{Kepler}, showed that acoustic modes are not detected in nearly 40 \% of solar-type stars \citep[e.g.][]{mathur_revisiting_2019}. This non-detection depends on the rotation and magnetic activity. Indeed, rotation locally modifies the convective properties \citep[e.g.][]{stevenson_turbulent_1979}, which in turn influence the stochastic excitation. In a previous work (Bessila et al., submitted to A\&A), we showed that uniform rotation can inhibit acoustic mode amplitudes. However, convective zones rotate differentially: the difference in the rotation rate between mid-latitude and the equator can go up to 60\%, as highlighted in recent observational works \citep[e.g.][]{benomar_asteroseismic_2018}.

In the present study, we extend the theoretical formalism for the stochastic excitation of acoustic modes \citep{samadi_excitation_2001} to include the effects of differential rotation (DR), through its influence on convection. We use a Rotating Mixing-Length Theory (R-MLT) approach \citep{stevenson_turbulent_1979, augustson_model_2019} to model the local impact of DR on convection. We then estimate numerically the resulting influence on the power injected into the acoustic modes using the combination of the MESA stellar structure and evolution code \citep{paxton_modules_2011, paxton_modules_2013, paxton_modules_2015, paxton_modules_2018, paxton_modules_2019, jermyn_modules_2023} and the GYRE stellar pulsation code \citep{townsend_gyre_2013}.

\section{Theoretical framework}

\subsection{Differential rotation}
 \label{sub:dr}
To model the stellar rotation profile, we chose a conical latitudinal DR profile: $\Omega (\theta) = \Omega_{p} + \Delta \Omega \sin^2(\theta)$, where $\Omega_{p}$ is the rotation rate at the equator and $\Delta \Omega \equiv \Omega_{eq} - \Omega_{p}$ is the difference in the rotation rate between the equator and the poles. If $\Delta \Omega > 0$, the rotation rate at the equator is higher than the one at the poles: it is a solar DR profile. On the contrary, if $\Delta \Omega < 0$, the rotation profile is antisolar.
Moreover, the DR rate is not independent of the mean rotation. Some studies find $ |\Delta \Omega|/\Omega \propto \mathcal{R}o_f^{p}$ for $\mathcal{R}o_f > 0.2$ where $\mathcal{R}o_f$ is the fluid Rossby number which compares the inertia of convection to the Coriolis acceleration effect and scales like $\mathcal{R}o_f \propto M_{\star}^{1.93 \pm 0.05}/\Omega_{\star}$, where $p$ is an exponent that is found to be between $2$ and $6$ depending on the author \citep[e.g.][]{saar_starspots_2010, brun_differential_2017}. An antisolar DR regime is seen in numerical simulations for high fluid Rossby number, the transition from an anti-solar to a solar rotation regime is being observed to occur around $\mathcal{R}o_f \sim 1$ \citep[e.g.][]{noraz_impact_2022}.

\subsection{Stochastic excitation}

Following the theoretical models by \cite{samadi_excitation_2001, belkacem_mode_2009}, the power injected into a given acoustic mode ($n,\ell, m$) is $\mathcal{P}=\frac{16}{15 I^2} \pi^3 \int_{\mathcal{V}} d^3 x_0 \rho_0^2\left| \frac{\mathrm{d} \xi_{r ; n, \ell}(r)}{\mathrm{~d} r}\right|^2 Y_{\ell,m} (\theta,\varphi) Y^{*}_{\ell,m}(\theta, \varphi) \hat{S}_R\left(r,\theta,\omega_0\right),$ where $I$ is the mode inertia, $\rho_0$ the mean density, $\xi_{r ; n, \ell}$ the radial component of the eigenfunction, $Y_{\ell,m}$ the spherical harmonic of degree $\ell, m$, and $Y^{*}_{\ell,m}$ its complex conjugate. $\hat{S}_R$ is the contribution of the Reynolds stresses: 
$\hat{S}_R\left(\omega_0\right)=\int \frac{d k}{k^2} E^2(k) \int d \omega \chi_k\left(\omega+\omega_0\right) \chi_k(\omega),$ where $E$ is the kinetic energy spectrum and $\chi_k$ is the eddy-time correlation spectrum.

\subsection{Rotating convection}

To model rotating convection, we consider the framework of R-MLT developed by \cite{stevenson_turbulent_1979}, which is built on the assumption that convection is dominated by the linear mode that carries the most heat \citep{malkus_heat_1954}. This simple R-MLT model predicts that the convective velocity scales like $\Omega^{-1/5}$, diminishing when the rotation increases. It has successfully been compared to results from local direct numerical simulations \citep[see e.g.][]{barker_theory_2014}.

\section{Results}
\begin{figure}[ht!]
 \centering
 \includegraphics[width=0.5\textwidth,clip]{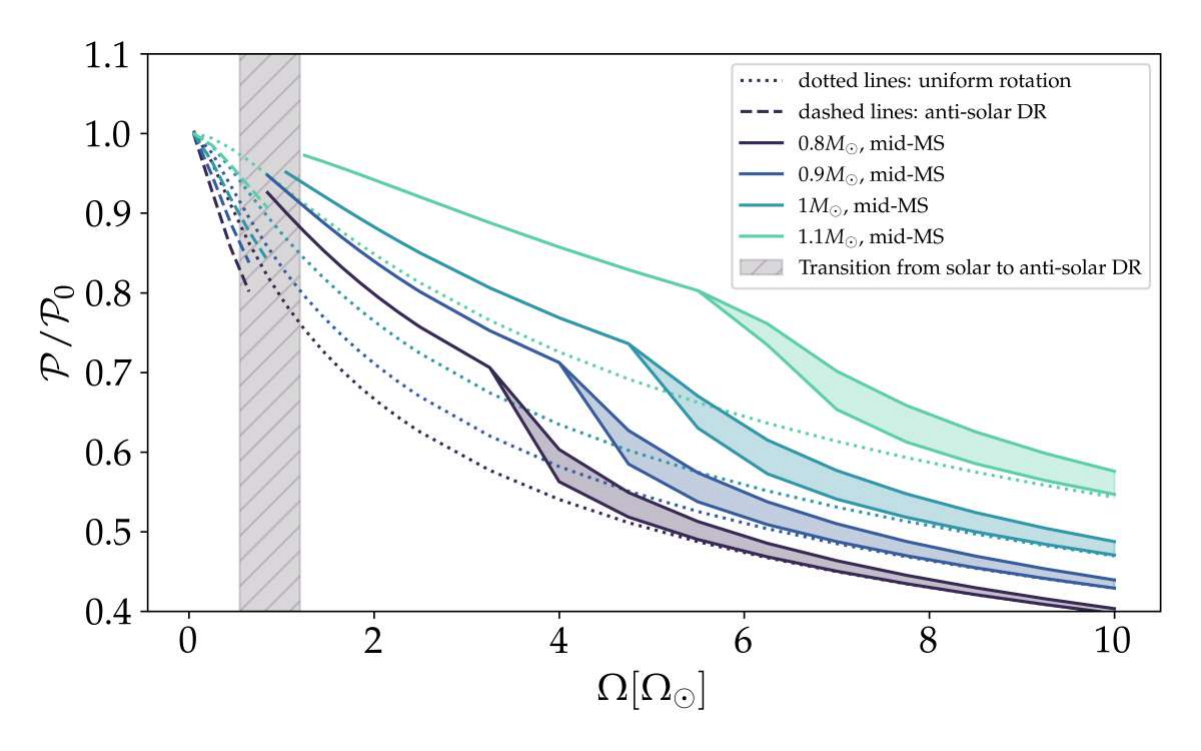}    
  \caption{Influence of DR on the power injected in the acoustic mode $\ell = 0, n=7, m=0$. The dashed lines denote the anti-solar rotation regimes  ($\mathcal{R}o_f > 1$), while the dotted lines show the uniform rotation case. We explored different possibilities for the exponent $p$ between 2 (upper limit) and 6 (lower limit). For values of $\mathcal{R}o_f > 0.2$, the relative differential rotation rate is assumed to be constant \citep[e.g.][]{brun_powering_2022} so the choice for $p$ has no impact.}
  \label{fig:rotdiff}
\end{figure}

 We examine the impact of DR on a fixed acoustic mode $\ell = 0$, $n=7$ in a main-sequence solar-like stars with masses ranging from $M = 0.8 M_{\odot}$ to $M = 1.1 M_{\odot}$, and metallicity $Z=0.02$. The stellar structure and oscillations are computed with the MESA and GYRE codes, respectively. At a given rotation rate $\Omega$, we constrain the DR using the scaling laws from Section \ref{sub:dr}. As shown in Fig. \ref{fig:rotdiff}, the power injected into the oscillations decreases more rapidly with an antisolar DR profile. With a solar DR profile, the resulting power diminishes less than in the solid body rotation case.  

\section{Conclusions}

We have extended the theoretical model for the stochastic excitation of acoustic modes in solar-like pulsators, taking into account a conical DR profile. We show that in stars with a lower rotation rate, which could display an antisolar DR, the power injected into the oscillations is lower than for the uniformly rotating case. The opposite occurs for solar DR. This result opens the way to a better characterisation of the modes detectability and exploration of DR in solar-like stars. 


\begin{acknowledgements}
G.B., L.B. and S.M. acknowledge support from the  European  Research Council  (ERC)  under the  Horizon  Europe programme  (Synergy  Grant agreement 101071505: 4D-STAR), from the CNES SOHO-GOLF and PLATO grants at CEA-DAp, and from PNPS (CNRS/INSU). While partially funded by the European Union, views and opinions expressed are however those of the author only and do not necessarily reflect those of the European Union or the European Research Council. Neither the European Union nor the granting authority can be held responsible for them.
\end{acknowledgements}

\bibliographystyle{aa}  
\bibliography{Biscarrat_S01.bib} 

\begin{thebibliography}{19}
\expandafter\ifx\csname natexlab\endcsname\relax\def\natexlab#1{#1}\fi

\bibitem[{Augustson \& Mathis(2019)}]{augustson_model_2019}
Augustson, K.~C. \& Mathis, S. 2019, The Astrophysical Journal, 874, 83

\bibitem[{Barker {et~al.}(2014)Barker, Dempsey, \& Lithwick}]{barker_theory_2014}
Barker, A.~J., Dempsey, A.~M., \& Lithwick, Y. 2014, The Astrophysical Journal, 791, 13, aDS Bibcode: 2014ApJ...791...13B

\bibitem[{Belkacem {et~al.}(2009)Belkacem, Mathis, Goupil, \& Samadi}]{belkacem_mode_2009}
Belkacem, K., Mathis, S., Goupil, M.~J., \& Samadi, R. 2009, Astronomy \& Astrophysics, 508, 345

\bibitem[{Benomar {et~al.}(2018)Benomar, Bazot, Nielsen, Gizon, Sekii, Takata, Hotta, Hanasoge, Sreenivasan, \& Christensen-Dalsgaard}]{benomar_asteroseismic_2018}
Benomar, O., Bazot, M., Nielsen, M.~B., {et~al.} 2018, Science, 361, 1231, aDS Bibcode: 2018Sci...361.1231B

\bibitem[{Brun {et~al.}(2022)Brun, Strugarek, Noraz, Perri, Varela, Augustson, Charbonneau, \& Toomre}]{brun_powering_2022}
Brun, A.~S., Strugarek, A., Noraz, Q., {et~al.} 2022, The Astrophysical Journal, 926, 21, aDS Bibcode: 2022ApJ...926...21B

\bibitem[{Brun {et~al.}(2017)Brun, Strugarek, Varela, Matt, Augustson, Emeriau, DoCao, Brown, \& Toomre}]{brun_differential_2017}
Brun, A.~S., Strugarek, A., Varela, J., {et~al.} 2017, The Astrophysical Journal, 836, 192, aDS Bibcode: 2017ApJ...836..192B

\bibitem[{Jermyn {et~al.}(2023)Jermyn, Bauer, Schwab, Farmer, Ball, Bellinger, Dotter, Joyce, Marchant, Mombarg, Wolf, Sunny~Wong, Cinquegrana, Farrell, Smolec, Thoul, Cantiello, Herwig, Toloza, Bildsten, Townsend, \& Timmes}]{jermyn_modules_2023}
Jermyn, A.~S., Bauer, E.~B., Schwab, J., {et~al.} 2023, The Astrophysical Journal Supplement Series, 265, 15, aDS Bibcode: 2023ApJS..265...15J

\bibitem[{Malkus(1954)}]{malkus_heat_1954}
Malkus, W. V.~R. 1954, Proceedings of the Royal Society of London Series A, 225, 196, aDS Bibcode: 1954RSPSA.225..196M

\bibitem[{Mathur {et~al.}(2019)Mathur, García, Bugnet, Santos, Santiago, \& Beck}]{mathur_revisiting_2019}
Mathur, S., García, R.~A., Bugnet, L., {et~al.} 2019, Frontiers in Astronomy and Space Sciences, 6

\bibitem[{Noraz {et~al.}(2022)Noraz, Brun, Strugarek, \& Depambour}]{noraz_impact_2022}
Noraz, Q., Brun, A.~S., Strugarek, A., \& Depambour, G. 2022, Astronomy \& Astrophysics, 658, A144

\bibitem[{Paxton {et~al.}(2011)Paxton, Bildsten, Dotter, Herwig, Lesaffre, \& Timmes}]{paxton_modules_2011}
Paxton, B., Bildsten, L., Dotter, A., {et~al.} 2011, The Astrophysical Journal Supplement Series, 192, 3, aDS Bibcode: 2011ApJS..192....3P

\bibitem[{Paxton {et~al.}(2013)Paxton, Cantiello, Arras, Bildsten, Brown, Dotter, Mankovich, Montgomery, Stello, Timmes, \& Townsend}]{paxton_modules_2013}
Paxton, B., Cantiello, M., Arras, P., {et~al.} 2013, The Astrophysical Journal Supplement Series, 208, 4, aDS Bibcode: 2013ApJS..208....4P

\bibitem[{Paxton {et~al.}(2015)Paxton, Marchant, Schwab, Bauer, Bildsten, Cantiello, Dessart, Farmer, Hu, Langer, Townsend, Townsley, \& Timmes}]{paxton_modules_2015}
Paxton, B., Marchant, P., Schwab, J., {et~al.} 2015, The Astrophysical Journal Supplement Series, 220, 15, aDS Bibcode: 2015ApJS..220...15P

\bibitem[{Paxton {et~al.}(2018)Paxton, Schwab, Bauer, Bildsten, Blinnikov, Duffell, Farmer, Goldberg, Marchant, Sorokina, Thoul, Townsend, \& Timmes}]{paxton_modules_2018}
Paxton, B., Schwab, J., Bauer, E.~B., {et~al.} 2018, The Astrophysical Journal Supplement Series, 234, 34, aDS Bibcode: 2018ApJS..234...34P

\bibitem[{Paxton {et~al.}(2019)Paxton, Smolec, Schwab, Gautschy, Bildsten, Cantiello, Dotter, Farmer, Goldberg, Jermyn, Kanbur, Marchant, Thoul, Townsend, Wolf, Zhang, \& Timmes}]{paxton_modules_2019}
Paxton, B., Smolec, R., Schwab, J., {et~al.} 2019, The Astrophysical Journal Supplement Series, 243, 10, aDS Bibcode: 2019ApJS..243...10P

\bibitem[{Saar(2010)}]{saar_starspots_2010}
Saar, S.~H. 2010, Proceedings of the International Astronomical Union, 6, 61

\bibitem[{Samadi \& Goupil(2001)}]{samadi_excitation_2001}
Samadi, R. \& Goupil, M.-J. 2001, Astronomy \& Astrophysics, 370, 136

\bibitem[{Stevenson(1979)}]{stevenson_turbulent_1979}
Stevenson, D.~J. 1979, Geophysical \& Astrophysical Fluid Dynamics, 12, 139, publisher: Taylor \& Francis

\bibitem[{Townsend \& Teitler(2013)}]{townsend_gyre_2013}
Townsend, R. H.~D. \& Teitler, S.~A. 2013, Monthly Notices of the Royal Astronomical Society, 435, 3406

\end{thebibliography}

\end{document}